\documentclass[pra,showpacs,twocolumn,amsmath,a4paper,preprintnumbers]{revtex4}
\usepackage{geometry}
\usepackage{bbm}
\usepackage{amssymb}
\usepackage{amsmath}
\usepackage{graphicx}
\usepackage{amsfonts}
\usepackage{hyperref}
\usepackage{oldgerm}
\usepackage{color}
\usepackage{epsfig}
\usepackage{amsthm}

\newcommand{\bra}[1]{\ensuremath{\left\langle #1\right|}}
\newcommand{\ket}[1]{\ensuremath{\left|#1\right\rangle}}
\newcommand{\braket}[2]{\ensuremath{\left\langle #1\vphantom{#2}\right.\left|\vphantom{#1}#2\right\rangle}}

\begin{document}

\title{Entanglement detection via mutually unbiased bases}
\author{Christoph Spengler$^1$}
\email{Christoph.Spengler@univie.ac.at}
\author{Marcus Huber$^{1,2}$}
\author{Stephen Brierley$^2$}
\author{Theodor Adaktylos$^1$}
\author{Beatrix C. Hiesmayr$^{1,3}$}
\affiliation{$^1$University of Vienna, Faculty of Physics, Boltzmanngasse 5, 1090 Vienna, Austria}
\affiliation{$^2$University of Bristol, Department of Mathematics, Bristol BS8 1TW, U.K.}
\affiliation{$^3$Masaryk University, Institute of Theoretical Physics and Astrophysics, Kotl\'a\v{r}sk\'a 2, 61137 Brno, Czech Republic}

\begin{abstract}
We investigate correlations among complementary observables. In particular, we show how to take advantage of mutually unbiased bases (MUBs) for the efficient detection of entanglement in arbitrarily high-dimensional, multipartite and continuous variable quantum systems. The introduced entanglement criteria are relatively easy to implement experimentally since they require only a few local measurement settings. In addition, we establish a link between the separability problem and the maximum number of mutually unbiased bases --- opening a new avenue in this long-standing open problem.
\end{abstract}


\preprint{UWThPh-2012-08}

\maketitle

\section{Introduction}
A key feature of quantum theory is the prediction of correlations that have no classical analogue, i.e. correlations that differ fundamentally from Bertlmann's socks \cite{Bertlmanns}. Whereas such quantum correlations were initially considered to be an artifact of the theory, it was later confirmed in several experiments that they actually exist in nature. They are a manifestation of the fact that composite quantum systems can be entangled, in the sense that they are not exclusively separable.

Nowadays, it is widely known that quantum entanglement enables numerous applications ranging from quantum cryptography to quantum computing. Although the theory of entanglement has been extensively studied within recent decades (for recent reviews consult Refs.~\cite{Horodeckireview,Guhnereview}), it is still an evolving research field with many open problems. One of these problems concerns the reliable and efficient detection of entanglement in experiments \cite{Altepeter,vanEnk}. While for bipartite two-level systems it is possible to experimentally verify the presence of entanglement by making a few joint local measurements, the number of measurements needed for entanglement detection generally scales rather disadvantageously with the size of the system. The main challenge for high-dimensional multipartite systems is not only to develop mathematical tools for entanglement detection, but to find schemes whose experimental implementation requires minimal effort. In other words, the aim is to verify entanglement with as few measurements as possible, specifically without resorting to full state tomography.

Another fundamental concept of quantum theory is complementarity, which states that there exist observables that cannot be measured simultaneously. In the mathematical formalism, complementarity expresses itself through the fact that there are pairs of observables for which no common eigenbasis can be found. Consequently, if two observables are complementary then it is impossible to prepare a system such that the outcome of both is predictable with certainty. The extreme case of complementarity is when the eigenbases of two observables form a pair of mutually unbiased bases (MUBs) \cite{Schwinger}. This is when all (normalized) eigenvectors of one observable have the same overlap with all eigenvectors of the other observable. Thus, if a system is in an eigenstate of a particular basis, then the measurement result in a corresponding mutually unbiased basis is completely random.

The question of how many MUBs exist for a given Hilbert space has been a lively topic of research (see \cite{Durt} for a recent review). Although it is been known since $1989$ \cite{wootters} that for $\mathcal{H}=\mathbb{C}^d$ the number of MUBs is at most $d+1$ and that such a \emph{complete set} of MUBs exists whenever $d$ is a prime power, the maximal number of MUBs remains open for all other dimensions.  Even for the smallest non-prime power dimension $d=6$, the existence of a complete set remains an open problem and current numerical \cite{Butterley,Brierley1,Raynal} and analytical \cite{Brierley2,Brierley3,jaming09,BruknerLat,Bengtssond6} evidence suggests that it is likely that there is none.

It is currently unclear if the (non-)existence of a complete set of MUBs in non-prime-power dimensions has fundamental reasons or consequences. However, one should also look at MUBs from a pragmatic perspective; or as phrased by Bengtsson \cite{Bengtssonthree}: ``...the real MUB problem is not how many MUBs we can find. The real MUB problem is to find out what we can do with those that exist.'' Existing applications of MUBs are quantum state tomography \cite{wootters,Filippov,Perez,Adamson}, cryptographic protocols \cite{Cerf,secretsharing} and the mean king's problem \cite{meanking1,meanking2}. In short, they are generally useful for finding and hiding (quantum) information.

In this paper, we present a new application of mutually unbiased bases. Namely, we link the concept of MUBs with the separability problem. We show that one can exploit the properties of MUBs to derive powerful entanglement detection criteria for arbitrarily high-dimensional systems. These criteria are well suited for the experimental verification of entanglement as they are experimentally accessible through measuring correlations between only a few local observables. In contrast to a full state tomography where the experimental effort can grow exponentially with the system size \cite{thew}, our approach enables optimal entanglement detection using a number of measurement settings which scales only linearly with the dimensionality of the local subsystems. In fact, we also show that even two local MUB settings, in general, suffice for a comparably robust entanglement test. Furthermore, by considering the noise thresholds of our criteria we find an interesting theoretical connection between the separability of density matrices and the maximum number of MUBs. In particular, we provide an alternative proof that there cannot be more than $d+1$ MUBs in any dimension. We also consider extensions of our methodology for continuous variables and multipartite systems. These are discussed by the example of the two-mode squeezed state and the Aharonov state.
\section{Preliminaries}
A set of orthonormal bases $\{\mathcal{B}_{k}\}$ for a Hilbert space $
\mathcal{H}=\mathbb{C}^{d}$ where $\mathcal{B}_{k}=\{|i_{k}\rangle
\}=\{|0_{k}\rangle ,\ldots ,|d-1_{k}\rangle \}$ is called mutually unbiased
(MU) iff
\begin{equation}
|\langle i_{k}|j_{l}\rangle |^{2}=\frac{1}{d}  \hspace{0.8cm}  \forall \ i,j \in \{0,\ldots,d-1\} \label{MU condition} \ ,
\end{equation}
holds for all basis vectors $|i_{k}\rangle $ and $|j_{l}\rangle $ that
belong to different bases, i.e. $\forall \ k\neq l$. If two bases are mutually unbiased, their corresponding observables are complementary --- a measurement of one of these observables reveals no information about the outcome of the other.

In dimension $d=2$, a set of three mutually unbiased bases is readily
obtained from the eigenvectors of the three Pauli matrices $\sigma
_{z},\sigma _{x}$ and $\sigma _{y}$:
\begin{align*}
\mathcal{B}_{1}&=\{ \ket{0_1}, \ket{1_1}\}=\{ \ket{0}, \ket{1}\} \ , \\
\mathcal{B}_{2}&=\{ \ket{0_2}, \ket{1_2}\}=\{ \frac{1}{\sqrt{2}} (\ket{0} + \ket{1}), \frac{1}{\sqrt{2}} (\ket{0} - \ket{1})\} \ ,\\
\mathcal{B}_{3}&=\{ \ket{0_3}, \ket{1_3}\}=\{ \frac{1}{\sqrt{2}} (\ket{0} + \mathbbm{i} \ket{1}), \frac{1}{\sqrt{2}} (  \ket{0} - \mathbbm{i} \ket{1})\} \ .
\end{align*}
These three bases constitute a complete set since it is impossible to find an additional basis that is mutually unbiased to all of them.

In general, for prime-power dimensions $d=p^{n}$, there are
several explicit methods to construct a complete set of $d+1$ MUBs making use
of finite fields \cite{wootters,KlappenRott}, the Heisenberg-Weyl group \cite{Bandyopadhyay}, generalized angular momentum
operators \cite{Kibler} and identities from number theory \cite{Archer}. For the special cases $d=2^n$ and $d=p^2$, it was shown that such sets can be constructed in a rather simple and experimentally accessible way \cite{Seyfarth,Wiesniak}.

The concept of mutually unbiased bases can also be extended to continuous variable (CV)
systems \cite{Durt,weigert08}. Here, the bases given by the (generalized) eigenstates of position and momentum operators
provide a well known example of MUBs. If one allows the right-hand-side of Eq.~(\ref{MU condition}) to vary between each pair of bases, a continuum of MUBs is available \cite{Durt}. Requiring that all pairwise overlaps have the same
modulus leads to a symmetric set of three MUBs for CV systems \cite{weigert08}.

First, in order to relate MUBs with the separability problem, let us specify how correlations can be quantified. Consider a bipartite system where measurements on each of the two subsystems $A$ and $B$ have $d$ different outcomes $\{0,\ldots,d-1\}$. If we can predict with certainty the outcome of a measurement on $A$ when we know the outcome of a measurement on $B$ (or vice versa) we call a system fully correlated. On the other hand, we call a system completely uncorrelated if the outcome of a measurement of one party tells us nothing about the other party, i.e. when the outcomes are completely random. Following this notion, it is possible to construct a correlation function for any two observables $a,b$ on $A,B$. We denote the joint probability that the outcome of $a$ is $i$ and the outcome of $b$ is $j$ by $P_{a,b}(i,j)$. We define the correlation function
\begin{align}
C_{a,b}=\sum_{i=0}^{d-1} P_{a,b}(i,i) \ ,
\end{align}
which we call the \emph{mutual predictability}. It can be used to quantify the probability of predicting the measurement results of $a$ knowing the outcome of $b$ and vice versa. Namely, if the observables $a$ and $b$ are fully correlated then the outcomes $\{i\}=\{0,\ldots,d-1\}$ can always be labeled in a way such that $C_{a,b}=1$. It is noteworthy that labels in general have no physical meaning. Thus, it is up to us what outcome we declare as $0,1,\ldots,\mbox{\emph{etc.}}$ . However, the point is that when the observables $a$ and $b$ are completely uncorrelated we obtain $C_{a,b}=1/d$ no matter what labeling we choose.

In the quantum case, each observable $a,b$ corresponds to an orthonormal basis $\{\ket{i_a}\}$ and $\{\ket{i_b}\}$. Here we have $P_{a,b}(i,j)=\bra{i_a}\otimes\bra{j_b} \rho  \ket{i_a}\otimes\ket{j_b}$ where $\rho$ is the state of the system, and thus the mutual predictability reads \mbox{$C_{a,b}= \sum_{i=0}^{d-1} \bra{i_a}\otimes\bra{i_b} \rho  \ket{i_a}\otimes\ket{i_b}$}. Again, one obtains $C_{a,b}=1$ for fully correlated states when $\{\ket{i_a}\}$ and $\{\ket{i_b}\}$ are chosen appropriately with respect to $\rho$, and $C_{a,b}=1/d$ for completely uncorrelated states, independent of the chosen bases.

\section{Entanglement detection: Bipartite qudit systems}
\label{biqudits}
For a particular state $\rho$ and measurement settings $a,b$ the quantity $C_{a,b}$ tells us nothing about the separability of a state. For instance, we can have $C_{a,b}=1$ for all entangled pure states $\ket{\psi}$ which directly follows from the Schmidt decomposition. Any entangled state may be written in the form $\ket{\psi}=\sum_{i=0}^{r} \lambda_ i \ket{i^s_a}\otimes\ket{i^s_b}$ with $1 \leq r \leq d-1$ using the orthonormal Schmidt bases $\{\ket{i^s_a}\}$ and $\{\ket{i^s_b}\}$. Using observables $a$ and $b$ that correspond to these bases, we obviously obtain $C_{a,b}=1$. However, we also obtain $C_{a,b}=1$ for a classically correlated separable state $\rho_{CC}=\sum_{i=0}^{r} |\lambda_ i|^2 \ket{i^s_a}\bra{i^s_a} \otimes \ket{i^s_b}\bra{i^s_b}$ as it yields the same joint probabilities $P_{a,b}(i,i)$ when we use $\{\ket{i^s_a}\}$ and $\{\ket{i^s_b}\}$.

Hence, to detect entanglement, the mutual predictability $C_{a,b}$ has to be measured in at least two bases, $a,b$ and $a',b'$. Let us consider a pure product state which we write as $\ket{\psi}_{\mbox{pro}}=\ket{0_1} \otimes \ket{0_1}$ in an arbitrary basis $\{\ket{i_1}\}$. For $\rho_{\mbox{pro}}=\ket{\psi}_{\mbox{pro}}\bra{\psi}_{\mbox{pro}}$ one obtains $C_{1,1}=1$ if both parties use the basis $\{\ket{i_1}\}$. However, in a second basis $\{\ket{i_2}\}$ which is mutually unbiased to $\{\ket{i_1}\}$, the mutual predictability $C_{2,2}$ is completely lost: Since $\{\ket{i_1}\}$ and $\{\ket{i_2}\}$ are mutually unbiased we have that
\begin{align}
P_{2,2}(i,i)&=\bra{i_2,i_2} \rho_{\mbox{pro}} \ket{i_2,i_2} \ ,\\
&=\braket{i_2,i_2}{0_1,0_1} \braket{0_1,0_1}{i_2,i_2} \ , \\
&=\underbrace{|\braket{i_2}{0_1}|^2}_{1/d} \cdot \underbrace{|\braket{i_2}{0_1}|^2}_{1/d} \ , \\
&=\frac{1}{d^2} ,
\end{align}
and consequently $C_{2,2}=\sum_{i=0}^{d-1}P_{2,2}(i,i)=1/d$.

Inspired by this result, let us consider the quantity $I_2=C_{1,1}+C_{2,2}$. As shown, with a pure product state we obviously can attain $I_2=1+\frac{1}{d}$ for a pair of MUBs. Similarly, we can achieve $I_m=\sum_{k=1}^{m} C_{k,k} = 1+ \frac{m-1}{d}$ for a product state using $m$ mutually unbiased bases $\mathcal{B}_{k}$ and corresponding terms $C_{k,k}$; because when the mutual predictability equals $1$ in one basis then it is $1/d$ with respect to the other $m-1$ bases. The main result of this paper is that these values are upper bounds for separable states, i.e. for all separable states and any set of $m$ mutually unbiased bases for $A$ and $B$ it holds that
\begin{align}
\label{maincriterion}
I_m=\sum_{k=1}^{m} C_{k,k} \leq 1+ \frac{m-1}{d} \ .
\end{align}
In particular, for a complete set of MUBs we have
\begin{align}
\label{maxcriterion}
I_{d+1}=\sum_{k=1}^{d+1} C_{k,k} \leq 2 \ .
\end{align}
\begin{proof}
For an arbitrary pure product state $|a\rangle \otimes |b\rangle \in \mathbb{C}^d \otimes \mathbb{C}^d$ we have
\begin{align}
I_m=\sum_{k=1}^{m}C_{k,k}=\sum_{k=1}^{m}\sum_{i=0}^{d-1}\left\vert \langle
i_{k}|a\rangle \right\vert ^{2}\left\vert \langle i_{k}|b\rangle \right\vert
^{2} \ .
\end{align}
Here, the inequality of arithmetic and geometric means $({x_1}+{x_2}+\ldots+x_n)/n \geq \sqrt[n]{x_1 \cdot x_2 \cdots x_n}$ for positive numbers implies that
\begin{align}
\sum_{k=1}^{m}C_{k,k}\leq \frac{1}{2}\sum_{k=1}^{m}\sum_{i=0}^{d-1}\left(\left\vert
\langle i_{k}|a\rangle \right\vert ^{4}+\left\vert \langle i_{k}|b\rangle
\right\vert ^{4} \right)\ . \label{AMGM}
\end{align}
Now we can exploit that for any pure state $\ket{a} \in \mathbb{C}^d$ and $m$ mutually unbiased bases it holds that
\begin{align}
\sum_{k=1}^{m}\sum_{i=0}^{d-1}\left\vert \langle i_{k} | a \rangle \right\vert
^{4}\leq 1+ \frac{m-1}{d} \ , \label{mstatement}
\end{align}
which was obtained in Ref.~\cite{Wu} as a generalization of the result established in Ref.~\cite{Larsen}. Thus, Eq.~(\ref{AMGM}) together with Eq.~(\ref{mstatement}) prove the validity of (\ref{maincriterion}) for all pure product states. Finally, since $I_m$ is linear in the density matrix $\rho$ it follows that (\ref{maincriterion}) holds for all (mixed) separable states as pure states represent extreme points.
\end{proof}

\begin{figure}[htb]
\centering
\includegraphics[width=8cm]{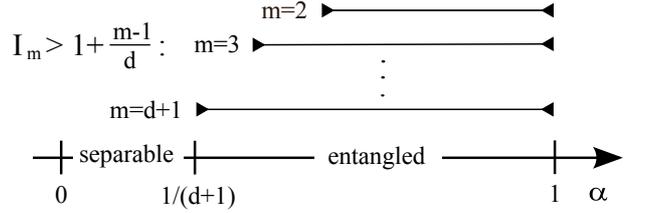}
\caption{Schematic illustration of the parameter regions detected by the criteria (\ref{maincriterion}) for the $d$-dimensional isotropic states $\rho_I=\alpha\ket{\phi^+_d}\bra{\phi^+_d}+\frac{1-\alpha}{d^2}\mathbbm{1}$ in dependence on the number of mutually unbiased bases $m$. The detection strength improves with increasing $m$ until for $m=d+1$ all entangled states are detected.}\label{isobound}
\end{figure}

The quantities $I_m$ together with the corresponding bounds for separable states can serve as criteria for entanglement detection in mixed states. However, what about the detection strength? Let us consider the $d$-dimensional isotropic states $\rho_I=\alpha\ket{\phi^+_d}\bra{\phi^+_d}+\frac{1-\alpha}{d^2}\mathbbm{1}$ with $\ket{\phi^+_d}=\frac{1}{\sqrt{d}}\sum_{i=0}^{d-1} \ket{i}\otimes\ket{i}$. These are known to be entangled for $\alpha>1/(d+1)$ and separable for $\alpha\leq1/(d+1)$ \cite{isobound}. For an arbitrary basis choice $x \leftrightarrow \{\ket{i_x}\}$ in system $A$ and $x^* \leftrightarrow \{\ket{i_x}^*\}$ in $B$ the mutual predictability is always $C_{x,x^*} = \alpha + (1-\alpha)/d$ since $\rho_I$ is $U\otimes U^*$ invariant \cite{isotropicstates}. Thus, using $m$ mutually unbiased bases $\{\mathcal{B}_k\}$ for $A$ and $\{\mathcal{B}^*_k\}$ for $B$ we attain $I_m=m(\alpha + (1-\alpha)/d)$ which violates (\ref{maincriterion}) for $\alpha>1/m$. Consequently, entanglement allows for values $I_m>1+ \frac{m-1}{d}$ which can be considered as an exact quantification of the statement that quantum correlations are more resistant against changes of the basis than ordinary correlations in separable states. As we also see the noise robustness of the criteria (\ref{maincriterion}) increases with the number of MUBs (see Figure~\ref{isobound}). If there exists a complete set of $m=d+1$ MUBs the criterion (\ref{maxcriterion}) is necessary and sufficient for the separability of $\rho_I$ as $\alpha=1/(d+1)$ is the exact bound of separability.

The significance of these results is manifold: First, the criteria (\ref{maincriterion}) are surprisingly powerful. Each isotropic state is local-unitarily equivalent to any other maximally entangled state mixed with white noise \cite{simplex}. By incorporating the corresponding local basis transformation that brings such a state into the isotropic form we can detect all entanglement when $d$ is of prime power dimension. Remarkably, only two MUBs are needed for detecting entanglement up to a threshold of $50\%$ noise. In comparison, Bell inequalities are often used as indicators of entanglement as they are simple to realize in experiments \cite{Guhnereview,Altepeter,vanEnk}. However, using two measurement settings for each party they merely reach a maximal noise threshold between $29.289\%$ and $32.656\%$ depending on the dimension $d$ \cite{CGLMP,Masanes}. Notably, two MUBs suffice to verify all entangled pure states in arbitrary dimension as is proven in Appendix \ref{sufficiency}. Moreover, regarding experimental verification of entanglement, we are now in the position that we can customize the number of MUBs depending on what is experimentally feasible.

Second, we emphasize that with the presented concept we establish a direct link between the separability boundary and the maximum number of MUBs (illustrated in Figure~\ref{isobound}). Notice that if there were $m>d+1$ MUBs for a Hilbert space $\mathcal{H}=\mathbb{C}^{d}$, then we would have $I_m>1+ \frac{m-1}{d}$ for separable states, namely for isotropic states $\rho_I$ with $1/m < \alpha\leq1/(d+1)$. This, however, is not compatible with the statement of Eq.~(\ref{maincriterion}), and thus we have shown by contradiction that there cannot exist more than $d+1$ MUBs.

Last, it should be noted that our criteria are adaptable for arbitrary mixed states, i.e. for verifying entanglement in density matrices beyond the white noise scenario. In general, if one applies our criteria to an arbitrary unclassified state $\rho$ one can improve the detection by maximizing the
outcome of $I_m$ over local-unitaries (by seeking the optimal transformation $\rho \rightarrow U_A \otimes U_B \rho U^{\dagger}_A \otimes U^{\dagger}_B$) and permuting the order of the basis vectors in the mutually unbiased bases. Appropriate tools for this optimization can be found in Refs.~\cite{SHHcp1,SHHcp2}. An analysis of a broader class of states which is related to a geometric structure of the Hilbert-Schmidt space is given in Appendix \ref{geometry}.

\section{Entanglement detection: Continuous variable states}
\label{CVS}
The concepts introduced in the previous section are not limited to discrete systems but can easily be applied to continuous variable (CV) states. As the noise robustness of the criteria (\ref{maincriterion}) increases with the number of MUBs, it is to expected that we can find quite strong entanglement detection criteria for CV systems since in this case there exist infinitely many MUBs \cite{Durt}. From a theoretical point of view it would certainly be interesting to study the generalization of our concept for a continuum of MUBs. However, in the current paper we take the viewpoint of an pragmatic experimentalist who has access to only a limited number of complementary observables. Let us study the simplest case where one has access to only two mutually unbiased bases corresponding to position $(x)$ and momentum $(p)$ measurements of single particles. Consider the two-mode squeezed state wave function \cite{Braunstein}
\begin{align}
&\psi_S(x_1,x_2)=\\
&\sqrt{\frac{2}{\pi}} \exp[-e^{-2r}(x_1+x_2)^2/2-e^{+2r}(x_1-x_2)^2/2] \ , \nonumber \\
&\psi_S(p_1,p_2)=\\
&\sqrt{\frac{2}{\pi}} \exp[-e^{-2r}(p_1-p_2)^2/2-e^{+2r}(p_1+p_2)^2/2] \ , \nonumber
\end{align}
depending on the squeezing parameter $r$, whose entanglement we would like to verify in an experiment by measuring joint probabilities. We use the mutual predictabilities $C_{x,x}=P_{x,x}(1,1)+P_{x,x}(2,2)$ of correlated positions
\begin{align}
P_{x,x}(1,1)&= \int_{-\infty}^{0} \int_{-\infty}^{0} |\psi_S(x_1,x_2)|^2 dx_1 dx_2 \ ,\\
P_{x,x}(2,2)&= \int_{0}^{\infty} \int_{0}^{\infty} |\psi_S(x_1,x_2)|^2 dx_1 dx_2 \ ,
\end{align}
and $C_{p,p}=P_{p,p}(1,2)+P_{p,p}(2,1)$ of anti-correlated momenta \footnote{Note that correlations and anti-correlations are the same up to the labeling of the measurement outcomes.}
\begin{align}
P_{p,p}(1,2)&= \int_{-\infty}^{0} \int_{0}^{\infty} |\psi_S(p_1,p_2)|^2 dp_1 dp_2 \ ,\\
P_{p,p}(2,1)&= \int_{0}^{\infty} \int_{-\infty}^{0} |\psi_S(p_1,p_2)|^2 dp_1 dp_2 \ .
\end{align}
Even though the correlations are measured quite imprecisely by dividing the state space into only two regions for each particle and observable (which can be regarded as a detector with very low resolution that produces only two distinguishable outcomes, equivalent to $d=2$) this suffices to detect almost all entanglement in a squeezed state: Via the minimal realization of our approach, i.e. $C_{x,x}+C_{p,p} \leq 1.5$ for separable states, we detect entanglement if the squeezing parameter is $r>0.3279$. This is already very close to the exact solution $r>0$ \cite{Braunstein}. Recall that this is done only by measuring correlations between positions $x_1,x_2$ and momenta $p_1,p_2$, that is without full knowledge of the state. Note that, if experimentally possible, we are always allowed to add further MUBs and use a finer partitioning of the Hilbert space (in accordance with the detector resolution) to improve the detection strength. However, in several cases few (or even only two) MUBs are enough to experimentally verify the presence of entanglement.

\section{Detection of genuine multipartite entanglement}
It is characteristic for multipartite systems that entanglement can occur in various ways. Here, it can happen that some parts of the system are entangled, while at the same time, others are separable \cite{Guhnereview,Horodeckireview,Gabrielksep}. For this reason, the concept of $k$-separability has been introduced: A pure state $\ket{\Psi}$ of an $n$-partite system is called $k$-separable if it can be written as a tensor product of $k$ vectors, i.e. $\ket{\Psi}=\ket{\psi_1}\otimes \cdots \otimes \ket{\psi_k}$. States that are $n$-separable do not contain any entanglement and are called \emph{fully separable}. Of special interest are quantum states whose entanglement ranges over all $n$ parties. Those are termed \emph{genuine multipartite entangled states} \cite{Guhnereview} and cannot be factorized at all, that is when $k=1$. The generalization to mixed states is straightforward: A mixed state $\rho$ is called $k$-separable if all pure state decompositions $\rho=\sum_i p_i \ket{\Psi_i}\bra{\Psi_i}$ require at least one $\ket{\Psi_i}$ which is at least $k$-separable according to the above definition.

While for pure states it is straightforward to examine if a state is genuine multipartite entangled, it is demanding to answer this question for mixed states. The main problem here is that standard entanglement criteria which are applicable to bipartite systems generally fail for the verification of genuine multipartite entanglement. This is due to the fact that biseparable states ($k=2$) can be entangled with respect to all bipartitions when they are mixed rather than pure: A typical example is a state of the form $\rho_{2\mbox{-sep}}=\frac{1}{3}(\rho_{A}\otimes\rho_{BC}+\rho_{B}\otimes\rho_{AC}+\rho_{C}\otimes\rho_{AB})$. Although this state is not genuine tripartite entangled, it might not be separable with respect to any fixed bipartition of the system.

Along with the fact that there currently exist only few criteria for the detection of genuine multipartite entanglement in mixed states (see e.g. Refs.~\cite{guehnecrit,HMGH,HESGH,Jungnitsch,spenglergme}) comes another problem to deal with. Namely, most of the currently known criteria are not scalable, that is in most cases the number of needed measurement settings grows exponentially with the number of parties. This is generally a serious obstacle to experimental implementations. In this section, we show that genuine multipartite entanglement can also be verified using few MUBs by adopting the previously introduced concept to the multi-particle scenario.

\begin{figure}[htb]
\centering
\includegraphics[width=8cm]{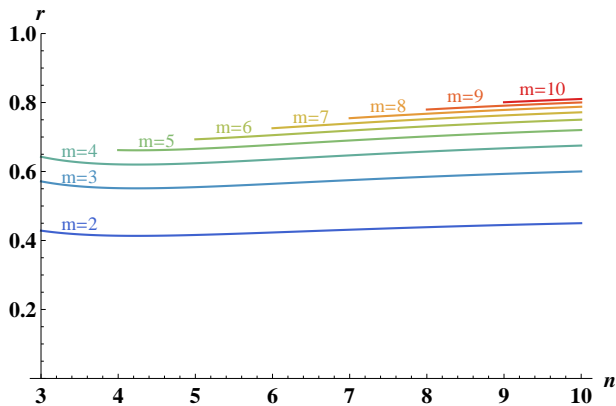}
\caption{(Color online) The noise robustness $r$ of the criteria (\ref{GMEcriterion}) for the n-partite Aharonov state in the presence of white noise, i.e. $\rho_{\mbox{aw}}=\alpha \ket{\mathcal{S}_n}\bra{\mathcal{S}_n} + \frac{1-\alpha}{n^n} \mathbbm{1}$. For $1-\alpha < r$ the state $\rho_{\mbox{aw}}$ is detected to be genuine multipartite entangled. The detection strength increases with the number of used mutually unbiased bases $m$.}\label{aharonovfig}
\end{figure}

Let us discuss our approach by the example of an $n$-partite $n$-dimensional singlet state \cite{cabello1,cabello2}, known as the \emph{Aharonov state} \cite{ahastate1,ahastate2}
\begin{align}
\ket{\mathcal{S}_n}=\frac{1}{\sqrt{n!}} \sum_{j,\ldots,l=0}^{n-1} \varepsilon_{j,\ldots,l} \ket{j, \ldots, l} \ ,
\end{align}
where $\varepsilon_{j,\ldots,l}$ denotes the generalized Levi-Civita symbol. For example, for three qutrits it reads
\begin{align}
\ket{\mathcal{S}_3}=\frac{1}{\sqrt{6}}(&\ket{012}+\ket{120}+\ket{201}\\
-&\ket{021}-\ket{102}-\ket{210}) \ . \nonumber
\end{align}
The Aharonov state has two central properties. First, from a correlation point of view, it is completely anti-correlated. This implies that if one performs measurements on $n-1$ parties and is aware of all outcomes then one can predict with certainty the outcome of the remaining party. Furthermore, this state is $U^{\otimes n}$ invariant implying that these anti-correlations always hold when all of the $n$ parties choose the same local basis \cite{cabello1,cabello2}. With respect to the mentioned symmetries of the state, it is reasonable to introduce an $n$-particle anti-correlation function
\begin{align}
A_{a,\ldots,z}&=\sum_{j,\ldots,l=0}^{n-1} | \varepsilon_{j,\ldots,l} | P_{a,\ldots,z}(j,\ldots,l)\\
 &= \sum_{j,\ldots,l=0}^{n-1} | \varepsilon_{j,\ldots,l} | \bra{j_a, \ldots, l_z} \rho  \ket{j_a, \ldots, l_z} \ ,
\end{align}
which is $A_{a,\ldots,z}=1$ iff all local measurement outcomes of the observables $\{a,\ldots,z\}$ are always unequal. Specifically, $A_{a,\ldots,z}=1$ for the Aharonov state when $a=\cdots=z$, i.e. when the same basis is chosen for all subsystems (as explained above). We build the linear combination
\begin{align}
\label{gmecriterion}
J_{m}=\sum_{a=1}^{m}A_{a,\ldots,a} \ ,
\end{align}
using $m$ mutually unbiased bases. This quantity $J_m$ is bounded by
\begin{align}
\label{GMEcriterion}
J_{m}\leq 1+\frac{m-1}{n} \ ,
\end{align}
for biseparable states.
\begin{proof}
Suppose we have a pure state $\ket{\Psi_{2\mbox{-sep}}}$ which is biseparable with respect to any bipartition $\{X|Y\}$. In general, such a state can reach $A_{a,\ldots,z}=1$ for a certain choice of observables $a,\ldots,z$. However, if we replace the local bases $\{a,\ldots,z\}$ by corresponding mutually unbiased bases $\{a,\ldots,z \} \rightarrow \{a',\ldots,z'\}$ then the predictability is lost, similarly to the bipartite qudit case (Sec.~\ref{biqudits}). We thus obtain $A_{a',\ldots,z'} \leq 1/\min \{d_X,d_Y\}$ where $d_X$ and $d_Y$ are the dimensions of $X$ and $Y$. Since $d=n$ is the minimum dimension over all bipartitions of the $n$-partite $n$-dimensional system it is guaranteed that $A_{a,\ldots,z} + A_{a',\ldots,z'} \leq 1 + 1/n$ holds for all biseparable states. Consequently, with $m$ MUBs we arrive at (\ref{GMEcriterion}), and since $J_m$ is linear in the density matrix $\rho$ it follows that any violation directly implies the existence of genuine multipartite entanglement in a (mixed) state.
\end{proof}

Let us discuss the detection strength of the criteria (\ref{GMEcriterion}) by the example of the Aharonov state in the presence of white noise $\rho_{\mbox{aw}}=\alpha \ket{\mathcal{S}_n}\bra{\mathcal{S}_n} + \frac{1-\alpha}{n^n} \mathbbm{1}$. For the pure Aharonov state $\ket{\mathcal{S}_n}$ we have $A_{a,\ldots,a}=1$ for all $a$, and for white noise $\frac{1}{n^n} \mathbbm{1}$ we have $A_{a,\ldots,a}=n!/n^n$. Thus, in total we obtain $J_{m}=m(\alpha + (1-\alpha)n!/n^n)$ which for
\begin{align}
 \alpha > \frac{n^n (m+n-1)-m n n!}{m n \left(n^n-n!\right)} \ ,
\end{align}
leads to a violation of $J_{m}\leq 1+\frac{m-1}{n}$. Figure~\ref{aharonovfig} illustrates the noise robustness of the criteria (\ref{GMEcriterion}) in dependence on the number of used mutually unbiased bases $m$. As can be seen therein, while the concept used is rather simple, the derived criterion is remarkably powerful in detecting genuine multipartite entanglement in the vicinity of the Aharonov state. For protocols where this particular state is used as a resource (e.g.~\cite{cabello1,Fitzi,cabello3}) this could be exploited to test whether the state was correctly distributed between all parties. Note that there currently exists no comparable test for verifying genuine multipartite entanglement in $\rho_{\mbox{aw}}$ and that the actual noise threshold is unknown. Note furthermore that it is to be presumed that our concept can easily be adopted to other multipartite states by taking into account their symmetries and correlations. In many cases this should lead to criteria with a valuable experimental-effort-to-detection-strength-ratio.
\section{Summary and Outlook}
In conclusion we have established a connection between mutually unbiased bases and entanglement detection. We showed that MUBs allow for an intuitive way of constructing entanglement criteria for arbitrarily high-dimensional systems. These criteria are beneficial for experiments since they require only a few local measurements. By means of the isotropic and Bell-diagonal states (Appendix~\ref{geometry}) we demonstrated that our approach can yield necessary and sufficient criteria for separability if a complete set of MUBs is available for the local subsystems. In addition, we found that the number of MUBs can be related to the separability problem and provided an alternative proof that for a $d$-dimensional system there cannot exist more than $d+1$ MUBs.

Besides optimal detection through complete sets of MUBs we showed that even using only two local complementary measurement settings it is possible to verify entanglement with a quite adequate robustness to noise. For experiments where the set of measurable observables is limited this may be of valuable help. For instance, for systems in high-energy physics investigated at accelerator facilities only a
restricted observable space is available due to the laborious effort and technical limitations. However, e.g. for neutral entangled $K$-mesons~\cite{Kaons} one could realize two MUBs during the time evolution of the system allowing for a direct test of entanglement via the introduced criteria. Two MUBs are also sufficient for detecting all entangled pure states of any two-qudit system (Appendix~\ref{sufficiency}) and allow for powerful entanglement detection in continuous variables. Even the presence of genuine multipartite entanglement can be tested very effectively through correlations in MUBs, which we demonstrated by the example of the Aharonov state.

For prime power dimensions, MUBs enable a complete state tomography. Consequently, local information and correlations with respect to MUBs should provide necessary and sufficient information to detect all entanglement in systems which are composed of subsystems with prime power dimensionality. For such systems, it should be possible to develop a general framework of entanglement detection based on complementary observables. For qubit systems, such a framework should be equivalent to the concept of correlation tensors (see e.g. Refs.~\cite{Paterek2,Paterek3,Vicente}), as the decomposition of density matrices in terms of Pauli matrices is intrinsically linked to MUBs. However, a generalization of correlation tensors to higher-dimensional systems has so far been addressed only by means of the generators of the special unitary group \cite{Vicente,Krammer,Yu}. Here, a theory in terms of MUBs should allow for an alternative method to investigate multilevel quantum correlations which is expected to be experimentally advantageous.

The presented scheme might also yield new results on systems with non-prime power dimensions: Just as we have shown that an upper bound on the number of MUBs can be deduced from the separability problem via the isotropic states, it might also be possible to determine the actual number of MUBs using a certain state and/or system. Finally, as numerous quantum features such as discord \cite{discord}, steering \cite{steering} and nonlocality (see Ref.~\cite{geometricNL} and references therein) give rise to particular correlations, it is conceivable that they can also be brought into relation with mutually unbiased bases, or even be directly formulated in terms of them.
\section*{ACKNOWLEDGEMENTS}
We would like to thank Andreas Winter, Renato Renner, Colin Wilmott, Shengjun Wu and Sergey Filippov for their valuable comments. CS and MH acknowledge financial support from the Austrian FWF (Project P21947N16) and the ERC. SB would like to thank the Heilbronn Institute for Mathematical Research for financial support. This project was co-financed by the SoMoPro programme. BCH acknowledges funding from the European Community within the Seventh Framework Programme under Grant Agreement No. 229603, and the COST action MP1006.

\appendix
\section{Sufficiency of two MUBs for pure states}
\label{sufficiency}
We show that two MUBs are sufficient to verify all entangled pure states of any bipartite qudit system $\mathcal{H}=\mathcal{H}_A\otimes\mathcal{H}_B=\mathbb{C}^d\otimes\mathbb{C}^d$. Assume our objective is to prepare a particular pure state $\ket{\Psi}=\sum_{m,n=0}^{d-1} c_{m,n} \ket{m}\otimes\ket{n}$. In order to achieve that the mutual predictability is maximal, i.e.
\begin{align}
C_{1,1}= \sum_{i=0}^{d-1} \bra{i_1}\otimes\braket{i_1}{\psi}\braket{\psi}{i_1}\otimes\ket{i_1}=1 \ ,
\end{align}
we use the measurement bases $\{\ket{i_1}\}$ on $A$ and $B$ for which our target state takes on the Schmidt form $\ket{\psi}=\sum_{i=0}^{r} \lambda_ i \ket{i_1}\otimes\ket{i_1}$ with $0 \leq r \leq d-1$, $\lambda_i \geq 0$ and $\sum_{i=0}^{r} \lambda_i^2=1$. For a second measurement of the mutual predictability $C_{2,2^*}$ we choose the (mutually unbiased) basis $\{\ket{i_2}\}=\{\ket{0_2},\ldots,\ket{{d-1}_2}\}$ with $\ket{i_2} = \frac{1}{\sqrt{d}} \sum_{k=0}^{d-1} \omega^{ki} \ket{k_1}$ and $\omega=\exp(2 \pi \mathbbm{i}/d)$, determined by the discrete Fourier transform. For the composite basis vectors we have
\begin{align}
\ket{i_2} \otimes \ket{i_2}^*= \frac{1}{d} \sum_{k,l=0}^{d-1} \omega^{(k-l)i} \ket{k_1} \otimes \ket{l_1} \ .
\end{align}
This leads to
\begin{align}
C_{2,2^*}= \sum_{i=0}^{d-1} & \bra{i_2}\otimes \bra{i_2}^* \ket{\psi}\braket{\psi}{i_2}\otimes\ket{i_2}^*\\
 = \sum_{i=0}^{d-1} & |\braket{\psi}{i_2}\otimes\ket{i_2}^*|^2 \ , \\
 =\sum_{i=0}^{d-1} & \left|\left[ \sum_{n=0}^{r} \lambda_n \bra{n_1}\otimes\bra{n_1} \right] \times  \right. \\
    \times & \left. \left[\frac{1}{d} \sum_{k,l=0}^{d-1} \omega^{(k-l)i} \ket{k_1,l_1} \right] \right|^2 \ . \nonumber
 \end{align}
We see that the only relevant vectors are those with $k=l$, in which case we have $\omega^{(k-l)i}=1$, and get
\begin{align}
C_{2,2^*}&= \sum_{i=0}^{d-1} | \frac{1}{d} \sum_{n=0}^{r} \lambda_n|^2 \ .
\end{align}
Here the squared absolute value $| \sum_{n=0}^{r} \lambda_n|^2$ can be rewritten as
\begin{align}
 C_{2,2^*}&= \frac{1}{d} (\underbrace{\sum_{n=0}^{r} \lambda_n^2}_{1}+ \sum_{m \neq n}^{r} \lambda_m \lambda_n) \ , \\
 &= \frac{1}{d} (1+ \sum_{m \neq n}^{r} \lambda_m \lambda_n) \ .
 \end{align}
Thus, altogether we obtain
\begin{align}
I_2=C_{1,1}+C_{2,2^*}= 1+ \frac{1}{d} (1+ \sum_{m \neq n}^{r} \lambda_m \lambda_n) \ .
\end{align}
For any separable state $\ket{\psi}$ the Schmidt rank is $1$, and consequently $\sum_{m \neq n}^{r} \lambda_m \lambda_n$ is zero since there is only one Schmidt coefficient $\lambda_m$ which equals $1$. Whereas, we have $\sum_{m \neq n}^{r} \lambda_m \lambda_n >0$ for any entangled state because they have Schmidt rank greater than or equal to $2$, i.e. there are at least two non-zero Schmidt coefficients $\lambda_m \geq 0$. Consequently, two MUBs are sufficient to detect all entangled pure states, as all of them achieve $I_2> 1+ \frac{1}{d}$. \ \footnote{Note that $I_2> 1+ \frac{1}{d}$ unambiguously implies the presence of entanglement regardless of which pairs of MUBs we use. However, just as for any entanglement verification scheme that does not require a full state tomography, we have to adjust our setup according to the expected state to achieve optimal detection.} \hfill $\square$

\section{Entanglement detection and geometry}
\label{geometry}
In Ref.~\cite{simplex}, a special simplex of locally maximally mixed two-qudit states,
also known as \emph{Bell-diagonal states}, was introduced. This set of states is given by
\begin{align}
\mathcal{W}=\{\sum_{k,l=0}^{d-1} c_{k,l}P_{k,l} \ | \ c_{k,l}\geq 0,\sum_{k,l=0}^{d-1}c_{k,l}=1 \} \ ,
\end{align}
where $P_{k,l}=\ket{\Omega_{k,l}}\bra{\Omega_{k,l}}$ are the
projectors of $d^2$ mutually orthogonal Bell states, generated by applying the unitary Weyl operators
\begin{align} W_{k,l}=\sum_{s=0}^{d-1}\omega^{sk}\ket{s}\bra{(s+l)\ \mbox{mod}\ d}
\end{align}
with $\omega=\exp(2 \pi \mathbbm{i}/d)$ and $k,l \in \left\{0,...,d-1\right\}$ on the maximally entangled
state $\ket{\Omega_{0,0}}=\frac{1}{\sqrt{d}}\sum_{i=0}^{d-1}\ket{i}\otimes\ket{i}$, i.e.
\begin{align}
\ket{\Omega_{k,l}}=\left(W_{k,l}\otimes\mathbbm{1}\right)\ket{\Omega_{0,0}} \ .
\end{align}
The
isotropic states from Sec.~\ref{biqudits} are also contained in this set. Here, for a complete set of MUBs, the quantity $I_{d+1}$ from Eq.~(\ref{maxcriterion}) reads
\begin{eqnarray}
I_{d+1}=1+ h d \ ,
\end{eqnarray}
where $h=\max\{c_{k,l}\}$ is the largest coefficient. Consequently, the region with $I_{d+1}\leq2$ corresponds to the so-called \emph{enclosure polytope} \cite{simplex}, whose facets are defined by the $d^2$ hyperplanes corresponding to optimal entanglement witnesses for all $\rho=\frac{1-\alpha}{d^2} \mathbbm{1}+\alpha P_{k,l}$. It was shown that all states outside this polytope are entangled \cite{simplex}. Hence, the quantity $I_{d+1}$ based on the maximum number of MUBs reflects the geometric structure of the enclosure polytope, which itself shares the symmetries of the simplex $\mathcal{W}$.


\begin{thebibliography}{99}

\bibitem{Bertlmanns}
J. S. Bell, J. Phys. Colloq. \textbf{42}, C2.41 (1981).

\bibitem{Guhnereview}
O. G\"uhne and G. Toth, Phys. Rep. \textbf{474}, 1 (2009).

\bibitem{Horodeckireview}
R. Horodecki, P. Horodecki, M. Horodecki, and K. Horodecki, Rev. Mod. Phys. \textbf{81}, 865 (2009).

\bibitem{Altepeter}
J. B. Altepeter, E. R. Jeffrey, P. G. Kwiat, S. Tanzilli, N. Gisin, and A. Ac\'{i}n, Phys. Rev. Lett. \textbf{95}, 033601 (2005).

\bibitem{vanEnk}
S. J. van Enk, N. L\"utkenhaus, and H. J. Kimble, Phys. Rev. A \textbf{75}, 052318 (2007).

\bibitem{Schwinger}J. Schwinger, Proc. Nat. Acad. Sci. U.S.A. \textbf{46}, 560 (1960).

\bibitem{Durt}
T. Durt, B.-G. Englert, I. Bengtsson, and K. \.{Z}yczkowski, Int. J. Quant. Inf. \textbf{8}, 535 (2010).

\bibitem{wootters}
W. K. Wootters and B. D. Fields, Ann. Phys. \textbf{191}, 363 (1989).

\bibitem{Butterley}
P. Butterley and W. Hall, Phys. Lett. A \textbf{369}, 5 (2007).

\bibitem{Brierley1}
S. Brierley and S. Weigert, Phys. Rev. A \textbf{78}, 042312 (2008).

\bibitem{Raynal}
P. Raynal, X. L\"u, and B.-G. Englert, Phys. Rev. A \textbf{83}, 062303 (2011).

\bibitem{Brierley2}
S. Brierley and S. Weigert, Phys. Rev. A \textbf{79}, 052316 (2009).

\bibitem{Brierley3}
S. Brierley and S. Weigert, J. Phys.: Conf. Ser. \textbf{254}, 012008 (2010).

\bibitem{jaming09}
P. Jaming, M. Matolcsi, P. M\'{o}ra, F. Sz\"{o}ll\H{o}si, and M. Weiner, J. Phys. A: Math. Theor. \textbf{42}, 245305 (2009).

\bibitem{BruknerLat}
T. Paterek, B. Dakic, and \v{C}. Brukner, Phys. Rev. A \textbf{79}, 012109 (2009).

\bibitem{Bengtssond6}
I. Bengtsson, W. Bruzda, \r{A}. Ericsson, J.-\r{A}. Larsson, W. Tadej, and Karol \.{Z}yczkowski, J. Math. Phys. \textbf{48}, 052106 (2007).

\bibitem{Bengtssonthree}
I. Bengtsson, AIP Conf. Proc. \textbf{889}, 40 (2007).

\bibitem{Filippov}
S. N. Filippov, V. I. Man'ko, Physica Scripta T \textbf{143}, 014010 (2011).

\bibitem{Perez}
A. Fern\'andez-P\'erez, A. B. Klimov, and C. Saavedra, Phys. Rev. A \textbf{83}, 052332 (2011).

\bibitem{Adamson}
R. B. A. Adamson and A. M. Steinberg, Phys. Rev. Lett. \textbf{105}, 030406 (2010).

\bibitem{Cerf}
N. J. Cerf, M. Bourennane, A. Karlsson, and N. Gisin, Phys. Rev. Lett. \textbf{88}, 127902 (2002).

\bibitem{secretsharing}
I.-C. Yu, F.-L. Lin, and C.-Y. Huang, Phys. Rev. A \textbf{78}, 012344 (2008).

\bibitem{meanking1}
B.-G. Englert and Y. Aharonov, Phys. Lett. A \textbf{284}, 1 (2001).

\bibitem{meanking2}
P. K. Aravind, Z. Naturforsch. \textbf{58a}, 85 (2003).

\bibitem{thew}
R. T. Thew, K. Nemoto, A. G. White, and W. J. Munro, Phys. Rev. A \textbf{66}, 012303 (2002).

\bibitem{KlappenRott}
A. Klappenecker and M. R\"otteler, Lect. Notes Comput. Sci. \textbf{2948}, 137 (2004).

\bibitem{Bandyopadhyay}
S. Bandyopadhyay, P. Boykin, V. Roychowdhury, and F. Vatan, Algorithmica \textbf{34}, 512 (2002).

\bibitem{Kibler}
M. Kibler and M. Planat, Int. J. Mod. Phys. B \textbf{20}, 1802 (2006).

\bibitem{Archer}
C. Archer, J. Math. Phys. \textbf{46}, 022106 (2005).

\bibitem{Seyfarth}
U. Seyfarth and K. S. Ranade, Phys. Rev. A \textbf{84}, 042327 (2011).

\bibitem{Wiesniak}
M. Wie\'sniak, T. Paterek, and A. Zeilinger, New J. Phys. \textbf{13}, 053047 (2011).

\bibitem{weigert08}
S. Weigert and M. Wilkinson, Phys. Rev. A \textbf{78}, 020303(R) (2008).

\bibitem{Wu}
S. Wu, S. Yu, and K. M\o lmer, Phys. Rev. A \textbf{79}, 022104 (2009).

\bibitem{Larsen}
U. Larsen, J. Phys. A: Math. Gen. \textbf{23}, 1041 (1990).

\bibitem{isobound}
R. A. Bertlmann, K. Durstberger, B. C. Hiesmayr, P. Krammer, Phys. Rev. A \textbf{72}, 052331 (2005).

\bibitem{isotropicstates}
M. Horodecki and P. Horodecki, Phys. Rev. A \textbf{59}, 4206 (1999).

\bibitem{simplex}
B. Baumgartner, B. C. Hiesmayr, and H. Narnhofer, J. Phys. A: Math. Theor. \textbf{40}, 7919 (2007).

\bibitem{CGLMP}
D. Collins, N. Gisin, N. Linden, S. Massar, and S. Popescu, Phys. Rev. Lett. \textbf{88}, 040404 (2002).

\bibitem{Masanes}
L. Masanes, Quant. Inf. Comp. \textbf{3}, 345 (2002).

\bibitem{SHHcp1}
C. Spengler, M. Huber, and B. C. Hiesmayr, J. Phys. A: Math. Theor. \textbf{43}, 385306 (2010).

\bibitem{SHHcp2}
C. Spengler, M. Huber, and B. C. Hiesmayr, J. Math. Phys. \textbf{53}, 013501 (2012).

\bibitem{Braunstein}
S. L. Braunstein and P. van Loock, Rev. Mod. Phys. \textbf{77}, 513 (2005).

\bibitem{Gabrielksep}
A. Gabriel, B. C. Hiesmayr, and M. Huber, Quant. Inf. Comp. \textbf{10}, 829 (2010).

\bibitem{guehnecrit}
O. G\"uhne and M. Seevinck, New J. Phys. {\bf 12}, 053002 (2010).

\bibitem{HMGH}
M. Huber, F. Mintert, A. Gabriel, and B. C. Hiesmayr, Phys. Rev. Lett. {\bf 104}, 210501 (2010).

\bibitem{HESGH}
M. Huber, P. Erker, H. Schimpf, A. Gabriel, and B. C. Hiesmayr, Phys. Rev. A {\bf 83}, 040301(R) (2011).

\bibitem{Jungnitsch}
B. Jungnitsch, T. Moroder, and O. G\"uhne, Phys. Rev. Lett. {\bf 106}, 190502 (2011).

\bibitem{spenglergme}
C. Spengler, M. Huber, A. Gabriel, and B. C. Hiesmayr, Quantum Inf. Process., DOI: 10.1007/s11128-012-0369-8 (arXiv:1106.5664)

\bibitem{cabello1}
A. Cabello, Phys. Rev. Lett. \textbf{89}, 100402 (2002).

\bibitem{cabello2}
A. Cabello, Phys. Rev. A \textbf{68}, 012304 (2003).

\bibitem{ahastate1}
M. Horodecki, P. Horodecki, R. Horodecki, J. Oppenheim, A. Sen(De), U. Sen, and B. Synak-Radtke, Phys. Rev. A \textbf{71}, 062307 (2005).

\bibitem{ahastate2}
L. Yuan and Z. Gui-Hua, Commun. Theor. Phys. \textbf{50}, 371 (2008).

\bibitem{Fitzi}
M. Fitzi, N. Gisin, and U. Maurer, Phys. Rev. Lett. \textbf{87}, 217901 (2001).

\bibitem{cabello3}
A. Cabello, J. Mod. Opt. \textbf{50}, 1049 (2003).

\bibitem{Kaons}
A. Di Domenico, A. Gabriel, B. C. Hiesmayr, F. Hipp, M. Huber, G. Krizek, K. M\"uhlbacher, S. Radic, C. Spengler, and L. Theussl, Found. Phys. \textbf{42}, 778 (2012).

\bibitem{Paterek2}
P. Badziag, \v{C}. Brukner, W. Laskowski, T. Paterek, and M. \.{Z}ukowski, Phys. Rev. Lett. \textbf{100}, 140403 (2008).

\bibitem{Paterek3}
W. Laskowski, M. Markiewicz, T. Paterek, and M. \.{Z}ukowski, Phys. Rev. A {\bf 84}, 062305 (2011).

\bibitem{Vicente}
J. I. de Vicente and M. Huber, Phys. Rev. A {\bf 84}, 062306 (2011).

\bibitem{Krammer}
R. A. Bertlmann and P. Krammer, J. Phys. A: Math.Theor. \textbf{41}, 235303 (2008).

\bibitem{Yu}
S. Yu and N.-L. Liu, Phys. Rev. Lett. \textbf{95}, 150504 (2005).

\bibitem{discord}
H. Ollivier and W. H. Zurek, Phys. Rev. Lett. {\bf 88}, 017901 (2001).

\bibitem{steering}
H. M. Wiseman, S. J. Jones, and A. C. Doherty, Phys. Rev. Lett. \textbf{98}, 140402 (2007).

\bibitem{geometricNL}
C. Spengler, M. Huber, B. C. Hiesmayr, J. Phys. A: Math. Theor. \textbf{44}, 065304 (2011).




\end{thebibliography}
\end{document}